\documentclass[12pt]{article}
\newcommand{\beq}{\begin{eqnarray}}
\newcommand{\eeq}{\end{eqnarray}}

\newcommand{\bi}{\bibitem}

\newcommand{\del}{\partial}

\newcommand{\bz}{{\bar z}}
\newcommand{\tphi}{{\tilde\phi}}

\def\be{\begin{equation}}
\def\ee{\end{equation}}
\def\bea{\begin{eqnarray}}
\def\eea{\end{eqnarray}}
\def\tr{{\rm\ Tr}}

\title{{\bf PP Wave Limit and Enhanced 
Supersymmetry in Gauge Theories}\\[2mm]}
\author{N. Itzhaki$^a$, Igor R. Klebanov$^a$ and Sunil
Mukhi$^{b,c,}$\footnote{On sabbatical leave from 
Tata Institute of Fundamental 
Research, 10/2001--9/2002.}\\ \\
\small \sl $^a$Department of Physics, Princeton University\\[-1mm]
         \small \sl Princeton, NJ 08544, USA\\
  \small \sl $^b$School of Natural Sciences, Institute for Advanced 
Study,\\[-1mm] 
\small \sl Princeton, NJ 08540, USA\\ 
\small\sl $^c$Tata Institute of Fundamental Research,\\[-1mm]
\small\sl Mumbai 400 005, India\\ \\
}

\begin{document}
\setlength{\baselineskip}{16pt}
\begin{titlepage}
\maketitle
\begin{picture}(0,0)(0,0)
\put(325,360){hep-th/0202153}
\put(325,345){PUPT--2024}
\put(325,330){TIFR/TH/02-06}
\end{picture}
\vspace{-36pt}
\begin{abstract}

We observe that the pp wave limit of $AdS_5\times M^5$
compactifications of type IIB string theory is universal, and
maximally supersymmetric, as long as $M^5$ is smooth and preserves
some supersymmetry. We investigate a specific case, $M^5=T^{1,1}$. The
dual ${\cal N}=1$ SCFT, describing D3-branes at a conifold
singularity, has operators that we identify with the oscillators of
the light-cone string in the universal pp wave background. The
correspondence is remarkable in that it relies on the exact spectrum
of anomalous dimensions in this CFT, along with the existence of
certain exceptional series
of operators whose dimensions are protected only
in the limit of large `t Hooft coupling. We also briefly examine the
singular case $M^5=S^5/Z_2$, for which the pp wave background becomes
a $Z_2$ orbifold of the maximally supersymmetric background by
reflection of 4 transverse coordinates.
We find operators in the corresponding
${\cal N}=2$ SCFT with the right properties to describe both
the untwisted and the twisted sectors of the closed string.

\end{abstract}
\thispagestyle{empty}
\setcounter{page}{0}
\end{titlepage}

\renewcommand{\baselinestretch}{1.2}  

\section{Introduction}

According to the AdS/CFT conjecture \cite{jthroat,US,EW}, the chiral 
operators
of the ${\cal N}=4$ supersymmetric $SU(N)$ gauge theory are in
one-to-one correspondence with the modes of type IIB supergravity
on $AdS_5 \times S^5$. 
The massive string modes, however, correspond to
operators in long multiplets whose dimensions diverge 
for large `t Hooft coupling as  $(g_{\rm YM}^2 N)^{1/4}$.
For this reason, the precise map between massive string modes and
gauge invariant operators has been difficult to construct at strong
coupling.
Recently, however, major progress in this direction has been made by
Berenstein, Maldacena and Nastase (BMN) \cite{BMN}.
Their proposal is to consider states with very large angular momentum
along the great circle of $S^5$, $ J\sim \sqrt N$. The metric felt
by such states is the Penrose limit of $AdS_5 \times S^5$, which is
\cite{papa1,papa2} the pp wave
\be\label{o}
ds^2 = -4 dx^+ dx^- + \sum_{i=1}^8  (d x_i)^2 
- \mu^2 (dx^+)^2 \sum_{i=1}^8  x_i^2
\label{ppwave}
\ee
supported by the 5-form RR field strength
\be
F_{+1234}= F_{+5678} \sim \mu
\ .
\ee
The 5-form breaks the $SO(8)$ symmetry of the metric down to
$SO(4)\times SO(4)$. The pp wave limit preserves 32 supercharges,
as many as the $AdS_5 \times S^5$ background \cite{papa1,papa2}.

This string background is remarkable in that the string theory
is exactly sovable in spite of the presence of the RR 5-form
field strength. As shown by Metsaev \cite{Metsaev} (see also
\cite{MT}),
in the light-cone gauge the 8 world sheet fields describing
the transverse coordinates, and their fermionic superpartners,
all acquire the same mass $\mu$. Therefore, the light-cone Hamiltonian
for a single string assumes the form
\be
p^- = \sum_{n=-\infty}^\infty N_n \sqrt {\mu^2 + {n^2\over 
(\alpha' p^+)^2 } }
\ ,
\ee
where $N_n$ is the excitation number of the $n$-th oscillator.

BMN combined this formula with the AdS/CFT duality to construct
the following relation between the dimension $\Delta$ and the
R-charge $J$ of the corresponding gauge theory operator \cite{BMN}:
\be
\Delta - J =\sum_{n=-\infty}^\infty N_n \sqrt {1 + {4\pi g_s N n^2\over
J^2} }
\ .
\ee
This formula is valid for $\frac{\Delta-J}{J} \ll 1$. 
In an ingenious construction, BMN identified the set
of single-trace operators which they argued 
have these dimensions and R-charges.
They suggested that this class of single-trace operators is 
in one-to-one correspondence
with the single string spectrum in the pp wave background
(\ref{ppwave}).

This remarkable result raises many interesting questions.
The questions we would like to ask are: Is the maximally supersymmetric
gauge theory necessary for reconstructing the full string 
spectrum? Can an analogous construction be carried out with a gauge
theory that has reduced supersymmetry or no supersymmetry at all?
In fact, in \cite{BMN} it was pointed out that the 
Penrose limit of the ${\cal N}=2$
supersymmetric orbifold $AdS_5\times S^5/Z_2$ is a $Z_2$
orbifold of the pp wave under 
$x_i\rightarrow - x_i$, $i= 5,6,7,8$.
In section 4 we consider the matching of string states in this 
background
with the single-trace operators of the ${\cal N}=2$
$SU(N)\times SU(N)$ orbifold gauge theory \cite{douglasmoore}.

Furthermore, we study the Penrose limits of spaces $AdS_5\times 
T^{p,q}$
which preserve ${\cal N}=1$ superconformal symmetry for $p=q$.
We will primarily consider the 
basic case $p=q=1$ which is dual to an ${\cal N}=1$ superconformal
$SU(N)\times SU(N)$ gauge theory \cite{KW}.
For $p=q>1$ we find a $Z_p$ orbifold of this gauge theory: an
${\cal N}=1$ superconformal
$SU(N)^{2p}$ gauge theory which we will not discuss here in any detail.
For $p\neq q$ the supergravity
background is non-supersymmetric and is, actually, unstable \cite{GM}.
Remarkably, for $p=q$ the Penrose limit is the maximally supersymmetric
pp wave background (\ref{ppwave}).\footnote {For $p\neq q$
we find a more general pp wave metric which gives,
in the light-cone gauge, different masses for the oscillators in the
$1234$, $56$ and $78$ directions.}
This is rather puzzling since the gauge theory of \cite{KW} has only
1/4 of the maximal supersymmetry. The only logical possibility seems to 
be
that the operators surviving in the appropriate limit of large R-charge
form a subsector with enhanced supersymmetry. While this is difficult
to prove, we will provide evidence that this is indeed the case.

Our construction of the string states largely parallels the BMN 
construction.
It will be very important for us that in addition to the $U(1)_R$
symmetry the gauge theory possesses $SU(2)\times SU(2)$ global 
symmetry.
The correct quantum number for classifying the string states 
(the analogue of $\Delta - J$ of \cite{BMN}) turns out to be
\be \label{lch}
H = \Delta - {J\over 2} + J_3 + J_3'
\ ,
\ee
where $J$ is the R-charge while $J_3$ and $J_3'$ are the $SU(2)$
and $SU(2)'$ quantum numbers. We find a family of $H=0$ 
operators: $\tr (A_2 B_2)^J$.\footnote{There also exist dibaryon 
operators \cite{GK}
carrying $H=0$: ${\rm det} A_2$ and ${\rm det} B_2$. These operators
are of no concern to our paper since they carry $J=N$, much greater
than in the BMN scaling for closed strings. These operators 
instead correspond
to wrapped D3-branes \cite{GK}. 
Such operators and their excitations may 
therefore be relevant to generalization of the BMN construction to
D-branes and open strings \cite{BHK}.}
 The insertion operators 
corresponding to the 8 transverse bosons and fermions carry
$H=1$ and are presented
in section 3. 

It is interesting that some of the operators we have to use are not 
chiral. For $J=0$ they have protected dimensions due
to the presence of the $SU(2)\times SU(2)$ global symmetry.
For $J>0$ it is likely that the dimensions
are not protected. However, we are able to use the AdS/CFT 
correspondence
to show that these operators have $H=1$ in the limit of very large
$g_s N$. This is in accord with the expectation that there is
a subsector of the gauge theory that has enhanced symmetry in
the appropriate limit of large quantum numbers and large
$g_s N$. 

\section{From ${\cal N}=1$ to ${\cal N}=4$ via the Penrose limit}

In this section we show that the Penrose limit of  $AdS_5 \times M^5$ 
is the 
same as for $AdS_5 \times S^5$ (that is, (\ref{o})) provided that $M^5$ 
is a
Sasaki-Einstein 5-manifold with a certain asymptotic behavior for the 
K\"ahler potential.
We apply this to $AdS_5 \times T^{11}$.

An $AdS_5\times M^5$ compactification of type IIB string theory, with
smooth $M^5$, preserves some supersymmetry if and only if $M^5$ is a
Sasaki-Einstein 5-manifold \cite{Acharya,Morrison}. This is equivalent
to saying that it is the base of a 6-dimensional cone $M^6$ with metric
\be
ds_{M^6}^2 = dr^2 + r^2 d\Omega_{M^5_{S-E}}^2
\ ,
\ee
where $M^6$ is K\"ahler and Ricci-flat, in other words a Calabi-Yau 
space. 

In turn, a Sasaki-Einstein 5-manifold $M^5_{S-E}$ is a $U(1)$
fibration over a 4-dimensional K\"ahler-Einstein space
$M^4_{K-E}$. Under the assumption that $M^5_{S-E}$ is smooth, its
metric can be written as follows \cite{Friedrich}:
\be
ds_{M^5_{S-E}}^2 = \left(d\beta + {i\over 2}
(K,_i dz^i- K,_{\bar i}d\bz^i)\right)^2
+ K,_{i{\bar j}}\,dz^i d{\bar z}^j
\label{semetric}
\ ,
\ee
where $K(z,\bz)$ is the K\"ahler potential of $M^4_{K-E}$.

We introduce a constant $R$ and scale the coordinates $z^i,\bz^i$ so
that the K\"ahler potential $K$ depends only on $({z^i\over R},
{\bz^i\over R})$. 
For the Penrose limit, we are interested in the behavior of the metric 
when 
$R\rightarrow\infty$.
Requiring that in this limit
the K\"ahler potential goes like
\be\label{m}
K(z,\bz)
{\longrightarrow} {z\bz\over R^2}
\ ,
\ee
one finds that
\be
{i\over 2}(K,_i dz^i - K,_{\bar i}d\bz^i)
{\longrightarrow} 
{i\over 2R^2}(\bz^i dz^i - z^i d\bz^i)
\ .
\ee
Consider now the full metric of $AdS_5\times M^5_{S-E}$
\[
\frac{ds^2}{R^2} = -dt^2 \cosh ^2\rho +d\rho^2 + \sinh^2 d\Omega_3^2+ 
\left(d\beta + {i\over 2} (K,_i dz^i - K,_{\bar i}d\bz^i)\right)^2
+ K_{,i{\bar j}}\,dz^i d{\bar z}^j
\ .
\]
With the geometry defined in these terms, the Penrose limit is
obtained by scaling the AdS coordinate $\rho = r/R$ and then
simply taking $R\rightarrow \infty$, dropping all terms that vanish in
this limit. The metric above then becomes
\be
ds^2 = - R^2 dt^2 + dr^2 + r^2 d\Omega_3^2  + 
R^2 d\beta^2  + i d\beta (\bz^i dz^i - z^i d\bz^i)
+ dz^i d{\bar z}^i
\ .
\ee
Replacing
\be
z^i\rightarrow e^{i\beta} z^i,
\ee
and defining
\be
x^+ = \frac12  (t+\beta), \quad x^- = \frac12 R^2(t-\beta)
\ ,
\ee
the metric is brought to the form
\be
ds^2 = -4 dx^+ dx^- - ({\vec r}\,^2 + z^i\bz^i)(dx^+)^2
+ d{\vec r}\,^2 + dz^i d\bz^i,
\label{ppmetric}
\ee
which is identical to Eq.(\ref{o}). 

In the remainder of the paper we shall mainly focus on a  special case 
of a
Sasaki-Einstein 5-manifold: $T^{1,1}$, the base of the conifold, 
whose metric can be written as \cite{cd,KW}
\be
ds^2_{T^{1,1}} = \frac19 (d\psi +\cos \theta_1 d\phi_1 +
\cos \theta_2 d\phi_2)^2+ \frac16 (d\theta_1^2 +
\sin^2 \theta_1 d\phi_1^2 +d\theta_2^2 +\sin^2 
\theta_2 d\phi_2^2)
\label{tmetric}
\ .
\ee
This metric can be written in the   general Sasaki-Einstein form 
Eq.(\ref{semetric}) by choosing the K\"ahler
potential of the 4-dimensional base $P^1\times P^1$ to be
\be
K(z^i,\bz^i) = {2\over 3}\sum_{i=1}^2 
\ln \left(1+\frac32 {z^i\bz^i\over R^2}\right)
\ .
\ee
Since Eq.(\ref{m}) holds for this choice of the K\"ahler potential 
it follows from the discussion above that
the Penrose limit of $AdS_5\times T^{1,1}$ is the same as the Penrose limit
of $AdS_5\times S^5$.
Recalling that the original compactification breaks $\frac34$ of the
supersymmetries of type IIB, we see that in the Penrose limit this
supersymmetry breaking becomes invisible, and maximal supersymmetry is
restored. As we will see shortly, this has interesting consequences
for the relationship between the field theory and its dual string. 

In light of the enhancement of supersymmetry in the Penrose limit
one might suspect that SUSY is not required all.
Starting with a general non-supersymmetric $AdS_5\times 
M^5$
and taking the Penrose limit, would one still end up with Eq.(\ref{o})?
To show that this is not the case, we consider $AdS_5\times T^{pq}$.
The metric is \cite{cd,GM}
\[
ds^2=a^2(d\psi +p \cos \theta_1 d\phi_1 +
q \cos \theta_2 d\phi_2)^2+ b^2 (d\theta_1^2 +
\sin^2 \theta_1 d\phi_1^2) +c^2(d\theta_2^2 +\sin^2 
\theta_2 d\phi_2^2),
\]
where $a,b$ and $c$ are determined by $p,q$ and the $AdS_5$ radius of 
curvature. The precise relation between $a,b,c$ and $p,q$ and $R$ is 
not 
important for us and can be found, for example, in \cite{GM}.
$\psi$ has periodicity $4\pi $ and $\phi_1, \phi_2$ 
have periodicity $2\pi$.

To take the Penrose limit it is useful to find a convenient null 
geodesic.
This can be done by defining
\be\label{22}
\tilde \psi = \psi + p \phi_1 + q \phi_2\ ,\quad
x^+={1\over 2} (t+a \tilde \psi)\ , 
\quad x^-={R^2\over 2}(t-a \tilde\psi)\ .
\ee
We expand for small $\theta_1$ and $\theta_2$: 
\be
b\theta_1 = r_1/R\ ,\qquad c\theta_2 = r_2/R\ ,
\ee
and take the $R\rightarrow \infty$ limit.
After shifting the angular coordinates,
\be
\varphi_1 = \phi_1 - {a p\over 2 b^2} 
\left (x^+ - {x^-\over R^2} \right )\ ,\qquad 
\varphi_2 = \phi_2 - {a q\over 2 c^2} 
\left (x^+ - {x^-\over R^2} \right ) 
\ , 
\ee
we find the following pp wave metric:
\beq\label{nonsusy}
&& ds^2 = -4 dx^+ dx^- + \sum_{i=1}^4  d x_i^2 +
\sum_{i=1}^2 (d r_i^2 + r_i^2 d\varphi_i^2)
\nonumber \\ &&
- (dx^+) {}^2 \left( \sum_{i=1}^4 x_i^2+ \frac{a^2p^2}{4 b^4}
 r_1^2 + \frac{a^2q^2}{4 c^4} r_2^2\right).
\eeq
For $p=q=1$, we have $a=1/3$, $b^2=c^2=1/6$, so that we recover
the metric (\ref{ppmetric}). 

For $p\neq q$,
not all the bosons  have the same mass in the light-cone
gauge, but the solution still has the general pp wave
form discussed in \cite{papa1,papa2}. Thus, it preserves at least 16
supersymmetries.\footnote{We are grateful to Nakwoo Kim,
Neil Constable and Jose Figueroa-O'Farrill for pointing out an
erroneous statement about supersymmetry for $p\neq q$ in
an earlier version of this paper.} 
The tachyons present in the
$AdS_5\times T^{pq}$ compactification \cite{GM} did not
 survive the pp wave limit, in agreement with the supersymmetry
 of the solution.
Since the
fermion masses in the light-cone gauge
are determined solely by the 5-form, they are the same 
as in the supersymmetric case. Therefore, some of the fermion masses
do not match bosons and the pp wave is not maximally 
supersymmetric \cite{papa1,papa2}.


\section{Field theory description of enhanced SUSY}

In the previous section we saw that the Penrose limit of $AdS_5 
\times T^{11}$ is identical to the Penrose limit of $AdS_5 \times S^5$.
The AdS/CFT duality implies that this should have non-trivial 
consequences
for the dual field theories. 

As was explained beautifully by BMN, from the dual field theory point 
of view  
taking the Penrose limit means focusing on a particular sector of 
single-trace operators of the ${\cal N}=4$ SYM theory
while taking the 't Hooft coupling to infinity.
To be more precise we need to focus on 
``almost BPS'' operators with large R-charge, $J$,
which  scales like the  square-root of the 't Hooft coupling:
\be
\lambda = g_{YM}^2 N \rightarrow\infty, \;\;\;\;\;
 \frac{J^2}{\lambda}=\mbox{finite},
\;\;\;\;\; \Delta -J=\mbox{finite}.
\ee
These operators are the ones relevant for describing 
strings propagating in the pp wave background.
The demand that $\Delta -J$ remains finite follows from
keeping the light-cone Hamiltonian finite. 
For example, the light-cone vacuum is identified with
$\tr Z^J$, where $Z= \phi_1 + i \phi_2$ carries R-charge $1$
($\phi_a$ are the 6 adjoint scalars fields of the ${\cal N}=4$
SYM theory).
For a given $J$ this is the unique  state with $\Delta-J=0$.
The 8 transverse oscillations of the string correspond to inserting
$\phi_a$, $a=3,4,5,6$ and $D_k Z$, $k=1,2,3,4$, into the trace.
The ${\cal N}=4$ SYM theory also has 4 doublets of adjoint Weyl 
fermions;
the 8 fermionic oscillators correspond to inserting them into the 
trace.
It is important that the above exhausts the list of operators whose
single insertion into the trace produces an operator with
$\Delta -J=1$. By studying their multiple insertions, BMN argued
that the resulting single-trace operators are in one-to-one 
correspondence
with the single string spectrum in the pp wave background
(\ref{ppwave}).

The result of the previous section implies that, even though the field 
theory dual to string theory on $AdS_5 \times S^5$ is not the same as 
the
 field theory dual to $AdS_5 \times T^{11}$, that particular sector must 
be equivalent. 
The aim of this section is to explain how this comes about from the 
point of view of the field theory dual to $AdS_5 \times T^{11}$
which has only ${\cal N}=1$ supersymmetry.
The ${\cal N}=1$ superconformal field theory on the worldvolume of $N$
D3-branes at a conifold singularity was first constructed in
Ref. \cite{KW} and has been extensively studied in subsequent works. We
briefly review the essential features of this theory here. For details,
the reader is referred to Ref.\cite{KW}.

The theory has a gauge group $SU(N)\times SU(N)$, along with
bi-fundamental superfields $A_1,A_2$ and
$B_1,B_2$. The superfields are doublets of the first and second
factors, respectively, of a global $SU(2)\times SU(2)$ symmetry. There
is a $U(1)_R$-symmetry under which the chiral multiplet all have
charge $+{1\over 2}$, and a quartic superpotential
\be
W = \lambda \epsilon^{ik}\epsilon^{jl}A_i B_j A_k B_l
\ .
\ee
This superpotential is marginal at the conformal fixed point by virtue
of the fact that the $A_i,B_i$ acquire anomalous dimensions. Thus, in
the CFT each of these fundamental fields has dimension 
$\Delta={3\over 4}$.

The first step toward finding the relevant sector in the theory which
is dual to the pp wave is to determine the light-cone Hamiltonian,
$i\partial_{x^{+}}$, in terms of the field theory generators
\footnote{The factor of $2$ in $J$ is due to the periodicity of
$\psi$ being $4\pi$.}
\be
\Delta = i\partial_t\ ,\quad J= -2i\partial_{\psi}\ , \quad
J_3= i\partial_{\phi_1}\ ,\quad J_3' =
i\partial_{\phi_2}\ .
\ee
Here $J$ is the $U(1)_R$ charge, and $J_3,J_3'$
are the diagonal generators of the two factors in the global symmetry
group $SU(2)\times SU(2)$.
We write
\be \partial_{x^{+}} = {\partial t\over \partial x^+} \partial_t
+ {\partial \psi\over \partial x^+} \partial_\psi +
{\partial \phi_i\over \partial x^+} \partial_{\phi_i}
\ .
\ee
Using the relation between the 
 original $T^{1,1}$ coordinates in
(\ref{tmetric}) and the
coordinates where the metric assumes
the pp wave form (\ref{nonsusy}),
\be
t = x^+ + {x^-\over R^2}\ ,\quad \psi= x^+ - {x^-\over R^2}
- \varphi_1-\varphi_2\ , \nonumber
\ee
\be
\phi_i = \varphi_i + x^+ - {x^-\over R^2}\ ,
\ee
 we find 
 the light-cone Hamiltonian given in (\ref{lch}), i.e.
\be
2P^-= H =\Delta - {1\over 2} J + J_3 + J_3'\ .
\ee
Analogously, we find
\be 2 P^+ = {1\over R^2} \left (\Delta + {1\over 2} J - J_3 - J_3'
\right )\ .
\ee

In Tables 1 and 2, we make a list of the $H$ values of the various
fundamental fields in the SCFT. Then we will construct some
gauge-invariant operators and explain how they are to be identified
with the theory of strings on a pp wave background.
\begin{table}[t]
\begin{center}
\begin{tabular}[t]{c|c|c|c|c|c|p{1cm} c|c|c|c|c|c|}
~ & $\Delta$ & $J$ & $J_3$ & $J_3'$ & $H$ & & & $\Delta$ & $J$ & $J_3$ 
&
$J_3'$ & $H$ \\[1mm]
\cline{1-6}\cline{8-13} &&&&&&&&&&&& \\[-4mm]
$A_1$ & ${3\over 4}$ & ${1\over 2}$ &  ${1\over 2}$ &  0 & 1 
&& ${\overline A_1}$ & ${3\over 4}$ & $-{1\over 2}$ &  $-{1\over 2}$ &  
0 &
${1\over 2}$ \\[1mm]
\cline{1-6}\cline{8-13} &&&&&&&&&&&& \\[-4mm]
$A_2$ & ${3\over 4}$ & ${1\over 2}$ &  $-{1\over 2}$ &  0 & 0 
&& ${\overline A_2}$ & ${3\over 4}$ & $-{1\over 2}$ &  ${1\over 2}$ &  
0 &
${3\over 2}$ \\[1mm]
\cline{1-6}\cline{8-13} &&&&&&&&&&&& \\[-4mm]
$B_1$ & ${3\over 4}$ & ${1\over 2}$ &  0  &  ${1\over 2}$ & 1 
&& ${\overline B_1}$ & ${3\over 4}$ & $-{1\over 2}$ &  0  &  $-{1\over 
2}$ &
${1\over 2}$ \\[1mm]
\cline{1-6}\cline{8-13} &&&&&&&&&&&& \\[-4mm]
$B_2$ & ${3\over 4}$ & ${1\over 2}$ &  0  &  $-{1\over 2}$ & 0 
&& ${\overline B_2}$ & ${3\over 4}$ & $-{1\over 2}$ &  0  &  ${1\over 
2}$ &
${3\over 2}$ \\[1mm]
\cline{1-6}\cline{8-13}\cline{1-6}\cline{8-13} &&&&&&&&&&&& \\[-4mm]
$\chi_{A_1}$ & ${5\over 4}$ & $-{1\over 2}$ &  ${1\over 2}$ &  0 & 2 
&& ${\overline \chi_{A_1}}$ & ${5\over 4}$ & ${1\over 2}$ &  $-{1\over 
2}$ &
0 & ${1\over 2}$ \\[1mm]
\cline{1-6}\cline{8-13} &&&&&&&&&&&& \\[-4mm]
$\chi_{A_2}$ & ${5\over 4}$ & $-{1\over 2}$ &  $-{1\over 2}$ &  0 & 1 
&& ${\overline \chi_{A_2}}$ & ${5\over 4}$ & ${1\over 2}$ &  ${1\over 
2}$ &  0
& ${3\over 2}$ \\[1mm]
\cline{1-6}\cline{8-13} &&&&&&&&&&&& \\[-4mm]
$\chi_{B_1}$ & ${5\over 4}$ & $-{1\over 2}$ &  0 & ${1\over 2}$ & 2 
&& ${\overline \chi_{B_1}}$ & ${5\over 4}$ & ${1\over 2}$ &  0 & 
$-{1\over 2}$
& ${1\over 2}$ \\[1mm]
\cline{1-6}\cline{8-13} &&&&&&&&&&&& \\[-4mm]
$\chi_{B_2}$ & ${5\over 4}$ & $-{1\over 2}$ &  0 & $-{1\over 2}$ & 1 
&& ${\overline \chi_{B_2}}$ & ${5\over 4}$ & ${1\over 2}$ &  0 & 
${1\over 2}$
& ${3\over 2}$ \\[1mm]
\cline{1-6}\cline{8-13}\cline{1-6}\cline{8-13} &&&&&&&&&&&& \\[-4mm]
$\psi$ & ${3\over 2}$ & 1 &  0 &  0 & 1 
&& ${\overline \psi}$ & ${3\over 2}$ & $-1$ &  0 &  0 & 2 \\[1mm]
\cline{1-6}\cline{8-13} &&&&&&&&&&&& \\[-4mm]
$\tilde\psi$ & ${3\over 2}$ & 1 &  0 &  0 & 1 
&& ${\overline {\tilde\psi}}$ & ${3\over 2}$ & $-1$ &  0 &  0 & 2 
\\[1mm]
\cline{1-6}\cline{8-13}
\multicolumn{6}{c}{\vtop{\smallskip
\hbox{\small Table 1: Dimensions and charges for}
\smallskip
\hbox{~~~~~~~~~~~\small chiral fields and gauginos}}} 
& \multicolumn{7}{c}{~~~~~~~~~\vtop{\smallskip\hbox{\small 
Table 2: Dimensions and charges 
for}\smallskip\hbox{~~~~~~~~~~~\small complex conjugate fields}}} \\
\end{tabular}\\
\end{center}
\end{table}
In Table 1, $A_i, B_i$ refer to the scalar components of the chiral
superfields described above, and $\chi_{A_i},\chi_{B_i}$ are their
fermionic partners. $\psi$ and $\tilde\psi$ are the gauginos of the two
gauge groups. Table 2 has the dimensions and charges of the complex
conjugate fields.

{}From the tables, we see that the unique operator that should be 
identified with the light-cone vacuum  is $\tr\, (A_2 B_2)^J$ which has
$H=0$. 
These operators are analogous to  $\tr\,Z^J$ in
the maximally supersymmetric theory of Ref. \cite{BMN}. 
Next, we turn to operators with $H=1$.
 Let us first consider the case $J=1$.
The large $J$ case will be discussed shortly. From the table we find 
the
following bosonic chiral operators: $ \tr\, A_1 B_2, \tr\,
A_2 B_1, \partial_k(\tr\,A_2 B_2) $, $k=1,2,3,4$. 
These are 6 of the necessary
operators, which means that we are missing two bosonic operators. 

For the fermionic operators, we find: $ \tr\, \chi_{A_2} B_2,
\tr\, A_2 \chi_{B_2} $ with $J=1$. As the $\chi$ are 2-component Weyl
fermions, these make 4 operators altogether. However, only two of them
are the superpartners of a bosonic operator, namely the combination
${\rm \tr\,}\left(\chi_{A_2} B_2 + A_2 \chi_{B_2}\right)$. This is the
variation of ${\rm \tr\,}A_2 B_2$. 

So far,
we have found  6 chiral  bosons and 2 
chiral fermions with $H=1$.
 However, for the correspondence with the universal pp wave 
to work, we must find  2 additional
bosons and 6 additional fermions.
It turns out that there are precisely two non-chiral bosonic 
operators with $J=0$ that have $H=1$.
The fact that the operators are non-chiral should not come as a 
surprise.
This is also the case  with the ${\cal N}=4$ 
SYM theory when viewed as a ${\cal N}=1$ 
theory with three chiral superfields \cite{Ferrara}.
Consider, for example, the operators
\be\label{3}
Z^J V,\;\;\;\;Z^J{\overline V},
\ee
where $Z=\phi^1+i\phi^2$ and $V=\phi^3+i\phi^4$. From the 
${\cal N}=4$ point of view both are chiral with protected dimensions.
However only the first one is chiral in the ${\cal N}=1$ language.

Indeed, much like in (\ref{3}) the two extra operators with $H=1$  
involve the complex conjugate components
of the scalars, and can be written as 
\be\label{uu}
\tr\, A_2{\overline A_1}, \;\;\; \tr\, B_2{\overline B_1}.
\ee 
Unlike in (\ref{3}) we do not have ${\cal N}=4$ supersymmetry 
at our disposal to fix their dimension.
The total dimension of these operators is not the naive sum of the
dimensions of the constituents. In general we would not know how to 
compute 
their dimensions, but it turns out \cite{ceresole,GM} that they belong to the
same supermultiplet as the currents generating the global $SU(2)\times
SU(2)$ symmetry. Therefore, their dimension is equal to its
free-field value, namely $\Delta=2$.\footnote{The argument goes as 
follows.
The global symmetry requires that there is an $SU(2)$
triplet of conserved currents of dimension 3. 
The $J_3=-1$ component of this triplet
is $ A_2 \partial_\mu {\overline A}_1 - 
(\partial_\mu A_2 ) {\overline A}_1 + 
{\rm fermions} .$
The scalar $A_2 {\overline A}_1$ is related by the supersymmetry generators 
to this current and hence has dimension $2$.}
Using the assignment of
R-charge and the other global charges for the conjugate fields, 
which can be found in Table 2, we see that indeed these operators have 
$H=1$.

Let us turn now to the fermionic operators.
The bosonic operators
we described above have fermionic counterparts, whose dimension is
correspondingly protected. These are $\tr\, {\overline \chi_{A_1}}A_2$ 
and $\tr\, {\overline \chi_{B_1}}B_2 $, which
provide us 4 additional fermionic operators. So we are still missing 
2 more fermionic operators to complete the set of 8.
These  can be constructed  by making use of the
gauginos. Acccording to the analysis of Ref. \cite{ceresole}, the
operators $\tr\, \left(\psi (A_2 B_2)^J + {\tilde\psi} (B_2
A_2)^J\right)$ lie in short multiplets.  Moreover, from Table 1, we
see that they have $H=1$.  Just as for the fermions in the chiral
multiplets, here also we keep only the symmetric combination (under
the interchange of the two gauge group factors) which is a protected
operator. 
Thus, we have found the last 2 fermionic
operators of $H=1$, making up the collection of 8 operators that we
propose to identify with the fermionic oscillators of the light-cone
superstring in the pp wave background.

In light of this counting it is very natural to propose that the 
following 8 bosonic and
8 fermionic operators, 
\be
\begin{array}{ll}
{\rm Bosonic:} & \tr\, A_1 B_2 (A_2 B_2)^J, 
 \tr\,A_2 B_1  (A_2 B_2)^J,\\[2mm]
&\tr\,A_2{\overline A_1} (A_2B_2)^J, \tr\, B_2 {\overline B_1} (B_2 
A_2)^J
\\[2mm]
&~\partial_i\tr\,(A_2 B_2)^J,  \\[3mm] 
{\rm Fermionic:} & {\rm \tr\,}(\chi_{A_2} B_2 + A_2 \chi_{B_2})(A_2
B_2)^J, \\[2mm]
&\tr\, {\overline \chi_{A_1}}A_2 (B_2A_2)^J, 
\tr\, {\overline \chi_{B_1}}B_2 (A_2B_2)^J, \\[2mm]
&\tr\, \left(\psi (A_2 B_2)^J + {\tilde\psi} (B_2
A_2)^J\right),
\end{array}
\ .
\ee
are the relevant ones to describe the BMN sector.
To establish this one needs to show that
these operators have $H=1$ for all $J$.
This follows immediately for the chiral operators. 
However, as far as we can
see, there is no field theory argument that protects the non-chiral 
operators.
That is, the argument used above for 
operators (\ref{uu}) cannot be generalized to $J>0$.
Since the Penrose limit involves sending the 't Hooft coupling to 
infinity,
it is sufficient for us to show that the operator of the form
\be
O_{k}= \tr\, A_2 {\overline A_1} (A_2 B_2)^k,
\ 
\ee 
has $H=1$ in the region where the SUGRA approximation is valid.
The quantum numbers of  
$O_{k}$ are $J=k$, ${\bf J} =1+ (k/2)$,
${\bf J'} = k/2$ (by ${\bf J}$ we denote the angular momentum of 
$SU(2)$
and analogously for the $SU(2)'$).
According to the AdS/CFT correspondence,
\be
\Delta = 2 + \sqrt{ 4 + (mL)^2},\;\;\;
(m L)^2 = \lambda + 16 - 8\sqrt{\lambda+ 4}\ ,
\ee
where \cite{GM} 
\be
\lambda = 6[ {\bf J} ({\bf J} +1) + {\bf J'} ({\bf J'} +1) - J^2/8]
\ ,
\ee
is the value of the Laplacian on $T^{1,1}$. For the operator $O_{k}$ we 
find
\be
(mL)^2 = (3k/2)^2 - 4\ ,
\ee
so that $\Delta= 2 + 3k/2$, which in turn implies $H=1$.

The fact that there is no apparent field theory argument to protect
the 
dimension of these operators suggests that there are non-trivial
$\alpha^{'}$ corrections away from the supergravity approximation.
This fits with the fact that the symmetry should be enhanced 
only in the strict Penrose limit.

\section{$S^5/Z_2$ Theory: Operator Spectrum and Strings on Orbifolded
PP Waves}

The compactification of type IIB string theory on $AdS_5\times
S^5/Z_2$ is an example where the compact 5-manifold is not smooth, but
has a singular $S^1$ submanifold. This compactification is dual to the
theory on D3-branes transverse to a $Z_2$ ALE singularity, and the
world-volume theory is an ${\cal N}=2$ SCFT.

The pp wave limit of this geometry can be obtained very simply and one
finds that the metric is the universal one, as given in
Eq. (\ref{ppmetric}) above. However, the coordinates $z_a$,
$i=1,2$, are
identified under the $Z_2$ action $z^i\rightarrow -z^i$, and so there
is an ALE singularity that survives in the transverse space.
We believe, however, that in the pp wave case
there is little physical difference
between the untwisted and twisted sectors. This is because
the untwisted states
have have no continuous transverse momenta; they are essentially
localized near the orbifold plane by the metric so that transverse
excitations have gaps. The same is true for the twisted sector states.
We will see that the gauge theory operators dual to the untwisted and 
twisted states are also very similar.

Although the field content of the ${\cal N}=2$
$Z_2$ orbifold gauge theory is similar to that for
$AdS_5\times T^{1,1}$ (and the two field theories are related by a
massive perturbation, as argued in Ref. \cite{KW}), the theory is
closer to the maximally supersymmetric one in an important aspect: the
fundamental fields have no anomalous dimensions, so we have scalars
and fermions of canonical dimension $\Delta=1,{3\over 2}$
respectively. The R-symmetry group is $SU(2)\times U(1)$. The gauge
group is $SU(N)\times SU(N)$, and each gauge multiplet contains a
complex adjoint scalar, which we denote $\phi,\tphi$. In addition
there are bi-fundamental fields which can be represented as ${\cal
N}=1$ chiral multiplets $A_i$, $B_i$, though in ${\cal N}=2$ they of
course combine into hypermultiplets.

In this theory the adjoint scalars $\phi,\tphi$ are associated with
positions of the (fractional) D3-branes
within the orbifold fixed sixplane. We choose the $U(1)$ subgroup
of R-symmetry to act only on these fields, but not on the
hypermultiplets.
Hence we can form gauge-invariant operators $\tr\,
\phi^J$ and $\tr\,\tphi^J$ which have $\Delta-J=0$. Each of these
appears to be independently analogous to the operators $\tr\, Z^J$ of
Ref. \cite{BMN}, which at large $J$ are identified with the vacuum
state of the string. Thus we seem to have two distinct candidates for
the string vacuum. However, note that these two operators are
exchanged by the $Z_2$ that exchanges the two gauge group
factors. This is the same group as the orbifold $Z_2$, and it is 
therefore
natural to associate the symmetric combination $\tr\,(\phi^J
+\tphi^J)$ with the ground state in the untwisted sector of the string, 
while the
antisymmetric combination $\tr\,(\phi^J -\tphi^J)$ is associated with
the ground state in the twisted sector. More precisely, the
operator $\tr\,(\phi^J
+\tphi^J)$ corresponds to a graviton moving with longitudinal momentum
$J$ while the operator $\tr\,(\phi^J -\tphi^J)$ describes a
member of the six-dimensional tensor multiplet moving with
longitudinal momentum $J$. More generally, we
define the operator $P$ which interchanges the two gauge groups,
and interchanges $\phi$ with $\tphi$, and $A_i$ with $B_i$,
$i=1,2$. Then, given any operator $O$, $O+ P O$ and $O-PO$
belong to the untwisted and twisted sectors respectively.

The operators  $\tr\,(\phi^J \pm \tphi^J)$
are rotated by the $U(1)$ factor in the
R-symmetry group $SU(2)\times U(1)$. The R-symmetry does not act on
the $A_i,B_j$, so each of these fields has $\Delta-J=1$. The new
feature, with respect to the maximally supersymmetric case, is that we
cannot form gauge-invariant operators out of $A_i$ or $B_j$ alone
since they are bi-fundamentals. Nevertheless, we can write down the
following set of `excited' operators where these operators appear
in pairs: 
\beq &&
 O^n_{ij} = \sum_{l=0}^J e^{\pi i nl/J}
\tr (\phi^l A_i \tphi^{J-l} B_j)
\ , 
\nonumber \\ &&
{\overline O}^n_{ij} = \sum_{l=0}^J e^{\pi i nl/J}
\tr (\phi^l {\overline B}_i \tphi^{J-l} {\overline A}_j)
\ , \nonumber \\ &&
O^n_{Aij} = \sum_{l=0}^J e^{\pi i nl/J}
\tr (\phi^l A_i \tphi^{J-l} {\overline A}_j)
\ , \\ &&
O^n_{Bij} = \sum_{l=0}^J e^{\pi i nl/J}
\tr (\phi^l {\overline B}_i \tphi^{J-l} B_j)
\ .\nonumber
\eeq
We may form their untwisted and twisted combinations,
which are, respectively, symmetric and antisymmetric
under the $Z_2$ symmetry $P$ interchanging the two gauge groups. 
The resulting operators correspond to
$a^i_{-n/2} \tilde a^j_{-n/2}$ acting on the (un)twisted light-cone vacuum, 
i.e. they describe oscillations of the closed string in the
$z^i$ directions, which we recall were
orbifolded by a $Z_2$ action. Note that for odd $n$ they have
twisted boundary conditions. In the gauge theory operators
we indeed pick up $(-1)$ by moving $A_i$ or $B_j$ around the loop:
for example, the $\tr (\phi^J A_iB_j)$ and 
$\tr (\tphi^J B_j A_i)$ terms
in $O^n_{ij}$ have a relative minus sign.\footnote{
We thank D. Berenstein for pointing this out to us.}
For $n=0$ they should correspond to the massless string states, hence
their dimensions should be protected so that these operators
have $\Delta = J+2$. It may seem peculiar that
non-chiral operators such as $ {\overline O}^n_{ij}$ should be protected, but
we already noted a similar phenomenon in the ${\cal N}=4$ case:
many of the operators that are chiral from the ${\cal N}=4$ point 
of view do not
look chiral when written in terms of ${\cal N}=1$ 
superfields \cite{Ferrara}.

Discussion of the oscillations of the string along the ${\vec r}$
directions, which arise from $AdS_5$, proceeds by analogy
with \cite{BMN}. Since the orbifold group does not
act on these directions, the corresponding oscillators do not
have to appear in pairs, and there cannot be
twisted boundary conditions. Each such insertion should have unit 
light-cone
Hamiltonian as before, and to these we associate the 
four $\Delta=J+1$ operators
$\del_k \tr\,(\phi^J \pm \tphi^J)$, $k=1,2,3,4$,
where the $\pm$ signs are chosen for the
untwisted and twisted sectors respectively.
The superpartners of these excitations are the operators
\be
\tr (\psi \phi^J) \pm \tr (\tilde \psi \tphi^J)
\ .
\ee
where $\psi\ (\tilde \psi)$ 
is one of the four complex fermions from the ${\cal N}=2$
vector multiplet which are the adjoints of the first (second) $SU(N)$.

The discussion above, although somewhat sketchy,
shows that it is indeed possible to construct
all the string states, in a $Z_2$ orbifold of the pp wave by a 
reflection
of 4 transverse coordinates, out of the gauge invariant operators of
the ${\cal N}=2$ superconformal $SU(N)\times SU(N)$ gauge theory.

\section*{Acknowledgements}
We thank D. Berenstein,  
S. Cherkis, E. Gimon, S. Gubser, A. Hashimoto, B. Kol, J. Maldacena, 
L. Pando-Zayas, A. Polyakov, L. Rastelli, 
A. Sen, S. Sethi, K. Skenderis, J. Sonnenschein, M. Strassler,
E. Verlinde and E. Witten for useful discussions.
The work of NI and IRK is supported in part by the NSF Grant
PHY-9802484. The work of SM is supported in part by 
DOE grant DE-FG02-90ER40542 and by the Monell Foundation.

\end{document}